\documentclass[prb,showpacs,aps,twocolumn]{revtex4}
\usepackage{amsmath}
\usepackage{graphicx}
\usepackage{dcolumn}
\usepackage{amssymb}

\begin{document}

\title{Magnetic field dependence of the many-electron states in a magnetic quantum dot: The ferromagnetic-antiferromagnetic transition}
\author{Nga T. T. Nguyen}
\email{nga.nguyen@ua.ac.be}
\author{F. M. Peeters}
\email{francois.peeters@ua.ac.be} \affiliation{Departement Fysica,
Universiteit Antwerpen, Groenenborgerlaan 171, B-2020 Antwerpen,
Belgium}


\begin{abstract}

The electron-electron correlations in a many-electron ($N_e=1, 2,
..., 5$) quantum dot confined by a parabolic potential is
investigated in the presence of a single magnetic ion and a
perpendicular magnetic field. We obtained the energy spectrum and
calculated the addition energy which exhibits cusps as function of
the magnetic field. The vortex properties of the many-particle wave
function of the ground state are studied and for large magnetic
fields are related to composite fermions. The position of the
impurity influences strongly the spin pair correlation function when
the external field is large. In small applied magnetic field, the
spin exchange energy together with the Zeeman terms leads to a
ferromagnetic-antiferromagnetic(FM-AFM) transition. When the
magnetic ion is shifted away from the center of the quantum dot a
remarkable re-entrant AFM-FM-AFM transition is found as function of
the strength of the Coulomb interaction. Thermodynamic quantities as
the heat capacity, the magnetization, and the susceptibility are
also studied. Cusps in the energy levels show up as peaks in the
heat capacity and the susceptibility.

\end{abstract}


\maketitle

\section{Introduction.}
Magnetic-doped quantum dots\cite{Furdyna} have attracted
considerable theoretical and experimental interests over the last
two decades. Diluted magnetic II-VI and III-V semiconductor (DMS)
quantum dots were fabricated with single-electron
control\cite{Klein,Chen}. A rich variety of different magnetic and
optical properties were
discovered\cite{Chang,Govorov,Archer,Besombes,Schmidt,Worjnar,Fernandez}.
In such structures one can explore the physical properties coming
from inter-carrier interactions and the interaction of the carriers
with the magnetic ion. This system promises to be relevant for
future quantum computing devices, where for instance the spin of the
magnetic ion is used as a quantum bit. More recently, electrically
active devices were fabricated in which a single manganese ion is
inserted into a single quantum dot \cite{Leger} with control of the
amount of charge in the quantum dot and consequently the possibility
of the control of magnetism of single Mn-doped quantum dots.

Investigation of the exact electronic structure of a two-dimensional
quantum dot confined by a parabolic potential containing several
electrons and a single magnetic impurity (in this paper $Mn^{2+}$)
in the presence of an applied magnetic field is a new topic. In a
recent investigation\cite{Cheng}, a three-dimensional (3D) Cd(Mn)Se
quantum dot containing several electrons, where only the low-energy
levels of the single-electron problem were taken into account, was
investigated in the presence of a magnetic field. Here we will
extend this work and include all relevant energy levels in order to
obtain a convergent solution for the ground state (and also excited
spectrum) of the system.

It is known that in the absence of a magnetic ion, an external
magnetic field is able to change the spin polarized state of weakly
interacting electrons $N_e$ in a quantum dot in such a way that in
the ground state it maximizes the total spin of the system: i.e.
$S=N_e/2$. If the inter-particle interaction is strong, even without
an applied magnetic field the electrons may already be polarized.
However, with increasing magnetic field and in the case the
inter-particle interaction is strong, the total spin of the system
can be unusually reduced by the magnetic field\cite{Marteen}.

In the present study, we investigate theoretically the few-electron
two-dimensional confined quantum dot system that contains a single
magnetic ion in the presence of an external magnetic field taking
into account a sufficient large number of single-particle orbitals
such that numerical ``exact" results are obtained. We explore how
sensitive the whole system is to the position of the magnetic ion in
the quantum dot and to the presence of a magnetic field and
investigate the competition between the following three energies: i)
the interaction of the magnetic ion with the electrons, ii) the
interaction of the magnetic ion with the magnetic field, and iii)
the interaction of the external field with the electrons. These
terms affect the spin polarization of the $N_e$ electrons in the
quantum dot.

Explicit studies of a $N_e$-correlated-electron system interacting
with a single magnetic ion in nonzero magnetic field are very rare
in the literature. Recent
theoretical\cite{Fernandez,Cheng,Hawrylak,NgaTTNguyen} and
experimental work\cite{Leger} has focused either on a small number
of electrons using the exact diagonalization approach at zero field
(Refs.\cite{Hawrylak,NgaTTNguyen}) for a $2$D quantum dot or at
nonzero field\cite{Cheng} including only the lowest single-particle
states for a 3D system or the exciton states relevant for optical
spectroscopy of self-assembled magnetic-doped quantum
dots\cite{Fernandez,Leger}.

Here, we will examine thoroughly the exact properties of the system
containing several correlated electrons and a single magnetic
impurity in the presence of a magnetic field. In our numerical
``exact" diagonalization approach we include an arbitrary number of
single-particle states to guarantee the accuracy of our results. We
investigate the influence of the strength of the inter-particle
interaction and the position of the magnetic ion on the ground state
of the system. We predict the interesting phenomenon that the
magnetic ion ferromagnetically couples with the electrons in a
region below a critical magnetic field and antiferromagnetically
with the electrons above this critical field. Thermodynamic
properties as magnetization, susceptibility, and the heat capacity
are investigated as function of magnetic field and temperature.
\begin{figure}[btp]
\begin{center}
\vspace*{-0.5cm}
\includegraphics*[width=9.2cm]{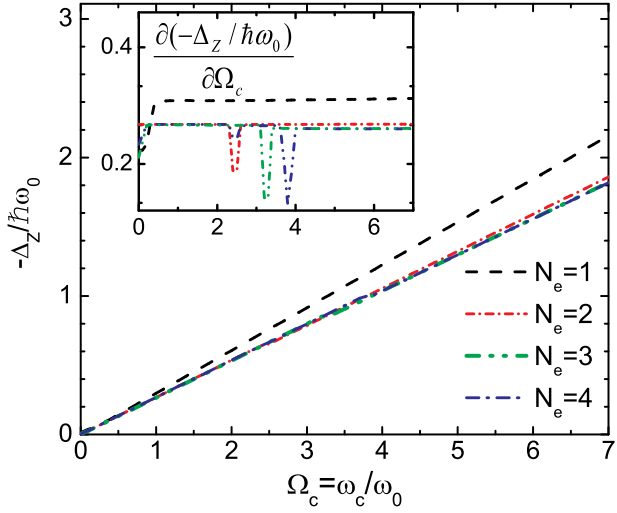}
\end{center}
\vspace{-0.5cm} \caption{(Color online) Total Zeeman energy for a
quantum dot with different number of electrons. The magnetic ion is
located at ($0.41l_0,0$), and the Coulomb interaction strength
$\lambda_C=0.2$. The inset is the first derivative of $\Delta_Z$
with respect to $\Omega_c$ that highlights the non-linearity of
$\Delta_Z$ in certain regions of the magnetic field.}
\label{localZeeman1234e}
\end{figure}

This paper is organized as follows. Section II introduces the model
and the numerical method. In section III, we present our numerical
results for the many-particle ground state and investigate
correlations through the appearance of vortices in the many-electron
wave function. Sect. IV addresses the many-particle spectrum and in
Sect. V we present results for different thermodynamic quantities.
Our discussion and conclusions are presented in Sect. VI.

\section{Theoretical model}
A quantum dot containing $N_e$ electrons with spins
$\overrightarrow{s_i}$ confined by a parabolic potential and
interacting with a single magnetic ion ($Mn^{2+}$) with spin
$\overrightarrow{M}$ and a magnetic field is described by the
following Hamiltonian:
\begin{widetext}
\begin{eqnarray} \label{e:Hamiltonian}
\hat{H} = \sum_{i=1}^{N_e}
 \left[\frac{1}{2m^*}{\left(-i\hbar\overrightarrow{\nabla}_{\overrightarrow{r}_i}+
    e\overrightarrow{A}(\overrightarrow{r_i})\right)^2}
    + \frac{1}{2}{m^*\omega_0^2\overrightarrow{r_i}^2}\right]+
\sum_{i<
 j=1}^{N_e}\frac{e^2}{4\pi\epsilon_0\epsilon|\overrightarrow{r_i}-\overrightarrow{r_j}|}
 \nonumber\\+
 \frac{1}{2}\hbar\omega_c\left(g_e m^*
 {S}_z+g_{Mn} m^*
 M_z\right)
 -J_c\sum_{i=1}^{N_e}\overrightarrow{s_i}\cdot
 \overrightarrow{M} \delta(\overrightarrow{r_i}-\overrightarrow{R}).
\end{eqnarray}
\end{widetext}
The vector potential $\overrightarrow{A}$ is taken in the symmetric
gauge: $\overrightarrow{A}=B/2(-y,x,0)$ where the magnetic field
$\overrightarrow{B}$ points perpendicular to the plane of the
interface. The confinement frequency $\omega_0$ is related to the
confinement length by: $l_0=\sqrt{{\hbar}/{m^*\omega_0}}$. $g_e$ and
$g_{Mn}$ are the Land\'{e} g-factor of the host semiconductor and
the magnetic ion, respectively. The dimensionless Coulomb strength
is defined as: $\lambda_C=l_0/a_B^{*}$ with
$a_B^{*}=4\pi\epsilon_0\epsilon\hbar^2/m^*e^2$ the effective Bohr
radius. The cyclotron frequency is: $\omega_c=eB/m^*$. $L_z$, $S_z$,
and $M_z$ are the projections of the total angular momentum of the
electrons, their spins, and the magnetic ion in the direction of the
magnetic field. The electrons and the magnetic impurity in the
quantum dot interact via the contact exchange interaction with
strength $J_c$.

We use the set of parameters\cite{Besombes,Hawrylak} that is
applicable to Cd(Mn)Te which is a II(Mn)VI quantum dot with typical
lateral size of about tens of nanometers. The dielectric constant
$\epsilon=10.6$, effective mass $m^*=0.106 m_0$, $a_B^{*}=52.9$
${\AA}$, $g_e=-1.67$, $g_{Mn}=2.02$, $J_c=1.5\times 10^{3} meV
{\AA}^{2}$, and $l_0$ about tens of nanometers ($\hbar\omega_0$
corresponding to tens of $meV$). For example, with
$\hbar\omega_0=51.32$ $meV$ gives $l_0=26.45$ {\AA}.

We rewrite the Hamiltonian in second-quantized form:
\begin{widetext}
\begin{eqnarray}\label{e:secondquantized}
\hat H =
\sum_{i,\sigma}E_{i,\sigma}c_{i,\sigma}^+c_{i,\sigma}+\frac{1}{2}\sum_{ijkl\sigma
\sigma^{'}} \langle i,j|V_0|k,l\rangle
c_{i,\sigma}^+c_{j,\sigma^{'}}^+c_{k,\sigma_{'}}c_{l,\sigma}
\nonumber +\frac{1}{2}\hbar\omega_c\left(g_e m^*
 {S}_z+g_{Mn} m^*{M}_z\right)
 \nonumber\\
 -\sum_{ij}^{}\frac{1}{2}{J_{ij}\left(\overrightarrow{R}\right)}
 \left[ \left(c_{i,\uparrow}^+
 c_{j,\uparrow}-c_{i,\downarrow}^+c_{j,\downarrow}\right)M_z +
 c_{i,\uparrow}^+c_{j,\downarrow}M^{-}
 + c_{i,\downarrow}^{+} c_{j,\uparrow} M^{+}\right],
\end{eqnarray}
\end{widetext}
where the first term is the single-particle energies $E_{i,\sigma}$
for an electron in state $i$ with spin $\sigma$ and the second term
is the Coulomb interaction. The third term is the electron and
magnetic ion Zeeman energy. The last sum is the electron-$\emph{Mn}$
interaction in which the first term describes the difference between
the number of spin up and spin down electrons, and the last two
terms describe the energy gained by flipping the electron spin along
side with flipping the spin of the magnetic ion. $M_z, M^{+}$, and
$M^{-}$ are the z-projection, raising and lowering operators,
respectively, of the magnetic ion spin (we consider $Mn$ ions which
have a spin of size $M$=5/2).

The single-particle states in a parabolic confinement potential
define a complete basis of Fock-Darwin orbitals
$\phi_{nl}\left(\overrightarrow{r}\right)$ and spin functions
$\chi_{\sigma}\left(\overrightarrow{s}\right)$:
\begin{equation}\label{e:basis}
\phi_{nls}\left(\overrightarrow{r},\overrightarrow{s}\right)=
\varphi_{nl}\left(\overrightarrow{r}\right)\chi_{\sigma}\left(\overrightarrow{s}\right),
\end{equation}
with the Fock-Darwin orbitals:
\begin{equation}\label{e:Fock-Darwin}
\varphi_{nl}\left(\overrightarrow{r}\right)=
\frac{1}{l_H}\sqrt{\frac{n!}{\pi\left(n+|l|\right)!}}\left(\frac{r}{l_H}\right)^{|l|}
e^{-il\theta}e^{-r^2/{2l_H^2}}L_n^{|l|}\left(r^2/l_H^2\right).
\end{equation}
In Hamiltonian (\ref{e:secondquantized}), $i$ denotes a set of
quantum numbers $\{n,l\}$ with $n,l$ the radial and azimuthal
quantum numbers, respectively. The effective length
$l_H=\sqrt{\hbar/m^*\omega_H}=l_0/\alpha$ in the presence of a
magnetic field is defined through the hybrid frequency
$\omega_H=\omega_0\sqrt{1+({{\omega_c}/{2\omega_0}})^2}$ where
$\alpha=\sqrt[4]{1+(\omega_c/2\omega_0)^2}$. The single-particle
orbital energy is given by:
\begin{equation}
E_{i,\sigma}=\hbar\omega_H (2n+|l|+1)-\hbar\omega_cl/2.
\end{equation}
The interaction parameters between the electrons and the magnetic
ion in the quantum dot is expressed:
\begin{equation}\label{e:JJex}
J_{ij}(\overrightarrow{R})=J_c
   \varphi_i^*(\overrightarrow{R})\varphi_j(\overrightarrow{R})
\end{equation}
as a product of two Fock-Darwin orbitals calculated at the position
of the magnetic ion.

We construct the many-particle wave function following the
configuration interaction (CI) method.
\begin{equation}
\Psi\left(\overrightarrow{x^*_1},\overrightarrow{x^*_2},...,\overrightarrow{x^*_{N_e}},
\overrightarrow{M}\right)=\sum^{N_{C}}_{k=1} C_k \Psi_{k},
\end{equation}
where $\Psi_k$ is the $k$-th state of the non-interacting many
electron wave function determined by $N_e$ electrons with $N_e$
different sets of quantum numbers $\left(n,l,\sigma\right)$ and the
single scatterer with one of the six states of the $Mn$.
$\overrightarrow{x^{*}_{i}}$=$\left(\overrightarrow{r_i},\overrightarrow{s_i}\right)$
stands for the coordinates and spin of a single electron. In second
quantization representation, the state $\Psi_k$, which is a Slater
determinant composed of single-electron states, can be translated
into a ket vector $|k\rangle$ grouping a total of $N_e$ electrons
into $N_{ \uparrow}$ electrons with z-component of spin up and
$N_{\downarrow}=N_e-N_{\uparrow}$ electrons with z-component of spin
down:
\begin{equation}
\Psi_k \Longrightarrow |k\rangle =
|c^+_{i_1\uparrow},c^+_{i_2\uparrow},...,c^+_{i_{N\uparrow}}\rangle|c^+_{j_1\downarrow},c^+_{j_2\downarrow},
...,c^+_{j_{N\downarrow}}\rangle|M_z\rangle.
\end{equation}
Here $i_{1\uparrow}\div i_{N\uparrow}$ and $j_{1\downarrow} \div
j_{N\downarrow}$ are the indices of the single-electron states for
which each index is a set of two quantum numbers (radial and
azimuthal quantum numbers), as mentioned above.

\begin{figure}[btp]
\begin{center}
\vspace*{-0.5cm}
\includegraphics*[width=9.0cm]{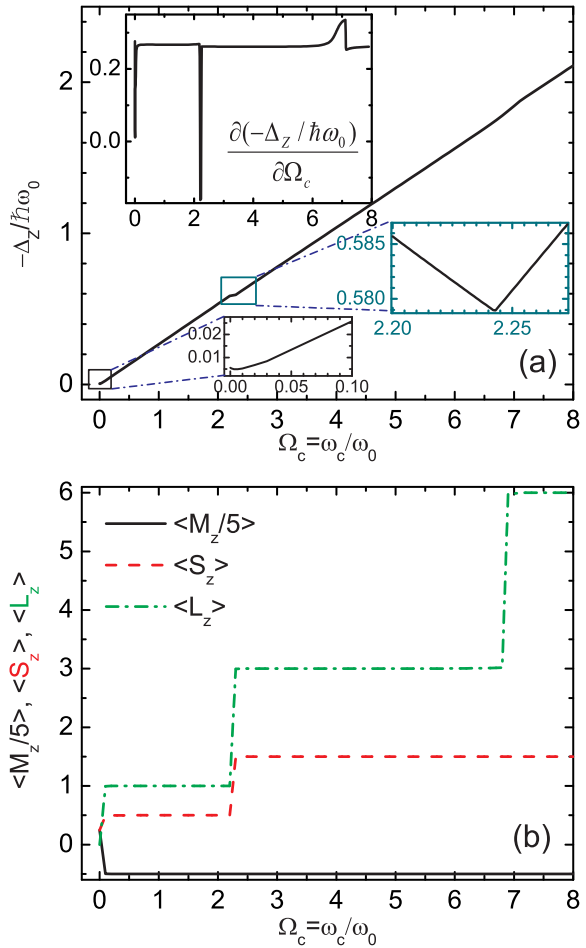}
\end{center}
\vspace{-0.5cm} \caption{(Color online) Total Zeeman energy (a)
calculated for a three-electron quantum dot with the magnetic ion at
($0.5l_0,0$) and $\lambda_C=0.5$. The upper inset is the first
derivative of the main plot with respect to $\Omega_c$ to visualize
more clearly the cusps. The lower inset is a zoom of the rectangular
regions of the main plot. (b) The z-projection of the total spin of
the magnetic impurity (solid black curve), of the electrons (dashed
red curve), and the total angular momentum (dash-dotted green
curve).} \label{Zeeman3e0.5l0ANDLzMzSz}
\end{figure}
The CI method, which is in principle exact if a sufficient number of
states are included, is limited to a small number of electrons due
to computational limitations. For a larger number of electrons
and/or magnetic ions, other approaches that e.g. are based on
spin-density functional theory (SDFT) using e.g. the local spin
density approximation (LSDA) as was used in Ref.\cite{Abolfath} is
able to handle a large number of electrons and/or magnetic ions. The
LSDA is exact only in the case of the homogeneous electron gas, and
in practice, works well also in most inhomogeneous systems. However,
in really highly correlated few-particle systems as discussed in
this paper, the LSDA might fail or be at least less accurate.
\begin{figure}[btp]
\begin{center}
\vspace*{-0.5cm}
\includegraphics*[width=9.2cm]{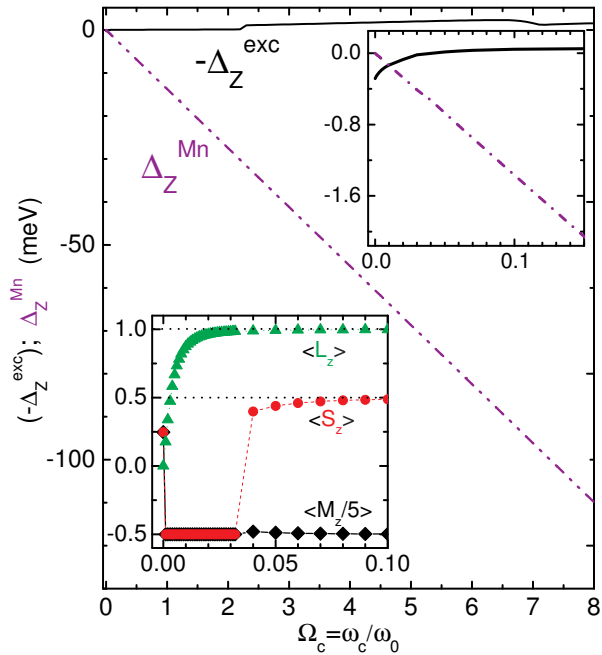}
\end{center}
\vspace{-0.5cm} \caption{(Color online) Zeeman energy of the
magnetic ion and exchange interaction of a three-electron quantum
dot with the magnetic ion located at $(0.5l_0,0)$ and the Coulomb
strength $\lambda_C=0.5$. The upper inset is a zoom of the small
magnetic field region and the lower one presents the average of
$M_z$, $S_z$, and $L_z$ for the ground state at very small magnetic
fields.}
 \label{twoterms_3e_0.5l0}
\end{figure}

\section{Ground-state properties}

\subsection{Zeeman energy}
We first explore the magnetic field dependence of the total Zeeman
energy:
\begin{equation}
\Delta_Z=E_{C}-E_{UC}=\Delta_Z^{electron}+ \Delta_Z^{Mn}+
(-\Delta_Z^{exc})
\end{equation}
which is the difference in energy in the presence and without a
magnetic ion. It consists of three terms: the difference of the
Zeeman energy of the electrons $\Delta_Z^{electron}$, the Zeeman
energy describing the interaction of the magnetic ion having spin
$\textbf{M}=5/2$ with the magnetic field, $\Delta_Z^{Mn}$, and the
exchange interaction of the ion with the electrons,
$-\Delta_Z^{exc}$. $\Delta_Z^{exc}$ is just the so called ``local
Zeeman splitting term" as discussed in Ref.\cite{NgaTTNguyen}. This
sum is basically the difference in the Zeeman energy of the
electrons between the cases with and without a magnetic ion plus the
energy contribution of the magnetic ion.
\begin{figure}[btp]
\begin{center}
\vspace*{-0.5cm}
\includegraphics*[width=9.0cm]{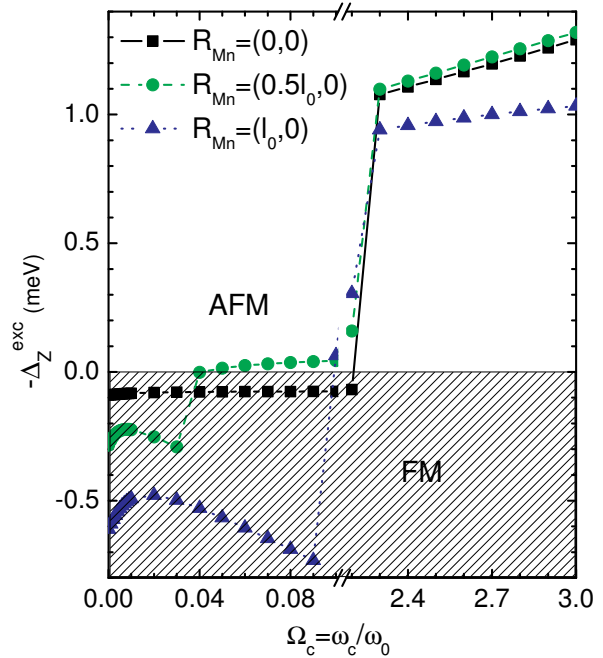}
\end{center}
\vspace{-0.5cm} \caption{(Color online) Exchange interaction of a
three-electron quantum dot with the magnetic ion located at three
different positions: $(0,0)$, $(0.5l_0,0)$, and $(l_0,0)$ for the
Coulomb strength $\lambda_C=0.5$. The horizontal line at
$\Delta_Z^{exc}=0$ separates the $\Delta_Z^{exc}$ plane into
ferromagnetic (FM) and antiferromagnetic (AFM) regions. }
 \label{exchange_3e_POS}
\end{figure}

For $N_e=1$, we find a total Zeeman energy that appears linear in
magnetic field. A similar linear behavior is also found in the cases
with $N_e>1$ but with different slopes (see
Fig.~\ref{localZeeman1234e}). Let us suppose, in Hamiltonian
(\ref{e:Hamiltonian}), that the contribution from the last term (the
local Zeeman energy or the exchange interaction term) is zero. For
instance this is the case when a magnetic ion is located at the
center of the quantum dot having three-electrons in the partially
filled $p$ shell, in which the first two-electrons fully fill the
$s$ shell and the remaining one is in either of the orbitals of the
$p$ shell. Then a perfect linear behavior of the total Zeeman energy
is found.

A closer look to $\Delta_Z$ gives us a slightly different picture,
as is provided by taking the derivative (see the inset of
Fig.~\ref{localZeeman1234e}). Notice that the total Zeeman term
$\Delta_Z$ has pronounced cusps and the number and positions of
these cusps is different for different number of electrons $N_e$.
There exists one at $\Omega_c=2.6$ for a two-electron quantum dot,
one at $\Omega_c=3.4$ for a three-electron quantum dot, and two at
$\Omega_c=2.5$ and $3.8$ for the four-electron quantum dot, with the
magnetic ion located at $(0.41 l_0, 0)$. The three-electron system
exhibits a much richer behavior when we increase the Coulomb
interaction strength to $\lambda_C=0.5$ as seen in
Fig.~\ref{Zeeman3e0.5l0ANDLzMzSz}(a) where we placed the magnetic
ion at $(0.5l_0,0)$. Cusps, which are highlighted in the two insets
in Fig.~\ref{Zeeman3e0.5l0ANDLzMzSz}(a), appear when the total
angular momentum and/or the total z-projection of the spin of the
electrons change abruptly with magnetic field. Notice that the total
Zeeman energy of a two-electron quantum dot in the presence of the
magnetic ion does not produce a similar behavior due to the fact
that the z-projection of the total spin is zero making the main
contribution (from the Zeeman spin term of the $Mn$-impurity)
negligible.
\begin{figure}[btp]
\begin{center}
\vspace*{-0.5cm}
\includegraphics*[width=9.0cm]{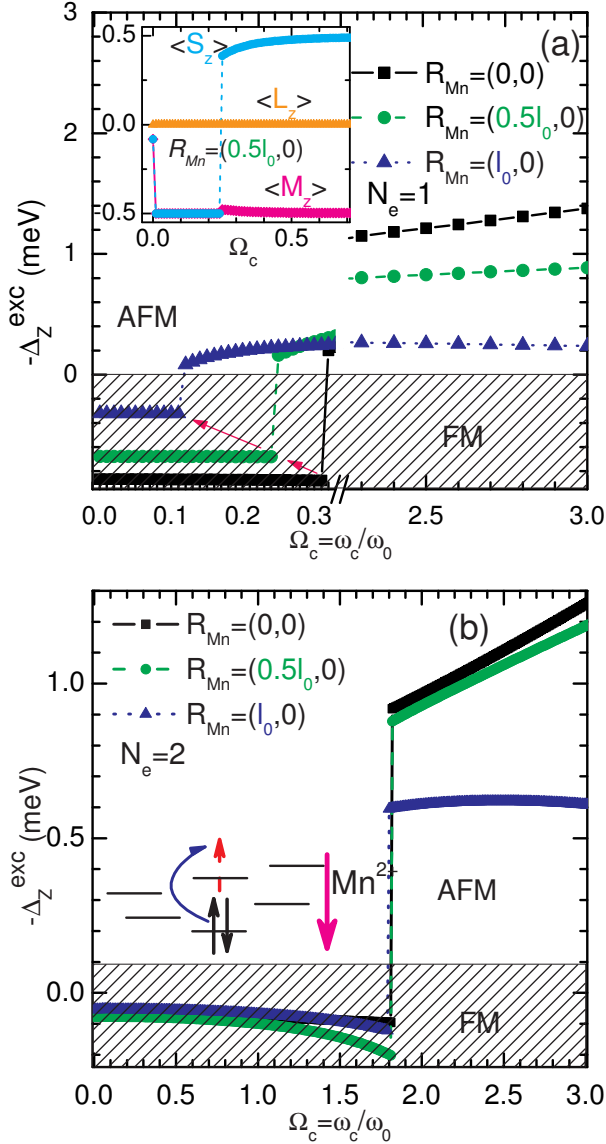}
\end{center}
\vspace{-0.5cm} \caption{(Color online) Exchange interaction of a
one-electron (a) and two-electron (b) quantum dot with the magnetic
ion located at three different positions and the Coulomb strength
$\lambda_C=0.5$. The inset in the upper plot are the averages of the
three quantities: $M_z$, $S_z$, and $L_z$ with the ion located at
$(0.5l_0,0)$ at small magnetic fields. The horizontal line separates
the plane into ferromagnetic (FM) and antiferromagnetic (AFM)
regions in both two plots. The schematic diagram in (b) explains why
the exchange energy is almost zero in the case $N_e=2$. }
 \label{exc1e2e}
\end{figure}

The Coulomb strength and the position of the magnetic impurity
affects the total Zeeman energy and influences the number and the
position of the cusps. The first pronounced cusp appears at lower
magnetic field for larger Coulomb interaction strength. This is a
consequence of the competition between the Coulomb energy and the
energy gap of the single-particle problem. Larger Coulomb strength
(smaller energy gap) leads to stronger electron-electron correlation
and consequently the electrons are more clearly separated from each
other. It results into a high probability for finding the electrons
to occupy higher energy states. That also means that the system
transfers to a configuration with larger $S_z$ and $L_z$ at smaller
applied field.
\begin{figure}[btp]
\begin{center}
\vspace*{-0.5cm}
\includegraphics*[width=9.0cm]{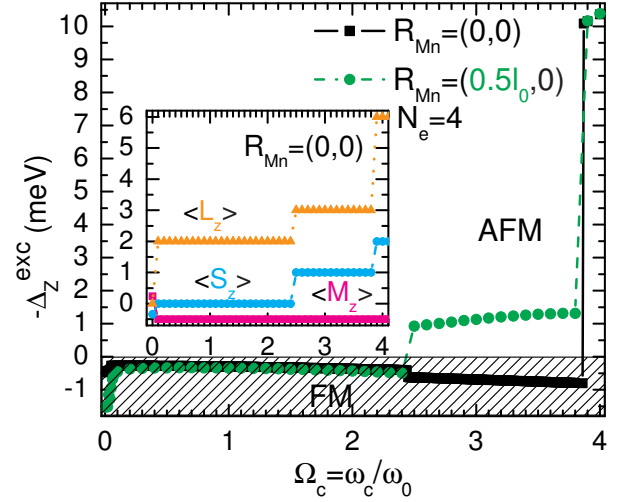}
\end{center}
\vspace{-0.5cm} \caption{(Color online) Exchange interaction of a
four-electron quantum dot with the magnetic ion located at three
different positions and the Coulomb strength $\lambda_C=0.2$. The
horizontal line separates the plane into ferromagnetic (FM) and
antiferromagnetic (AFM) regions. The inset shows the total
$z$-projection of the $M_z$, $S_z$, and $L_z$ of the electron system
with the magnetic ion located at the center of the dot.}
 \label{exc4e}
\end{figure}
\begin{figure}[btp]
\begin{center}
\vspace*{-0.5cm}
\includegraphics*[width=9.0cm]{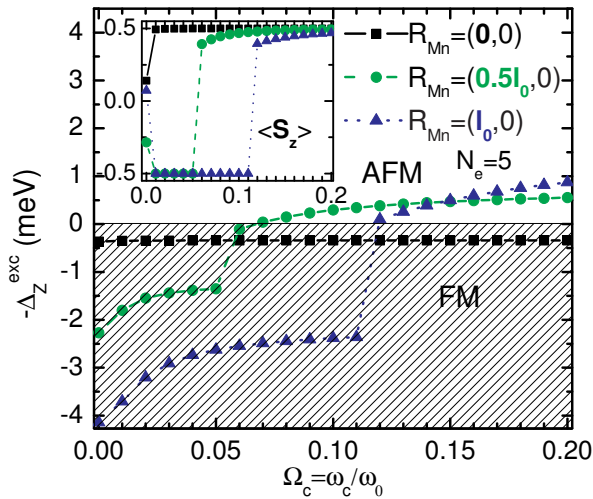}
\end{center}
\vspace{-0.5cm} \caption{(Color online) Exchange interaction of a
five-electron quantum dot with the magnetic ion located at several
positions and the Coulomb strength $\lambda_C=0.2$. The horizontal
line separates the plane into ferromagnetic (FM) and
antiferromagnetic (AFM) regions. The inset shows $S_z$ alone vs.
magnetic field.}
 \label{exc5e}
\end{figure}

We also found that the ground-state energy is sensitive to the
presence of the magnetic field. In zero magnetic field, the ground
state receives contributions from many different configurations
having z-projection of the magnetic ion, $M_z$, from $-5/2$ to
$5/2$. When a magnetic field is applied, the ground state favors
states with projection of the spin of the magnetic ion down and the
states with $M_z=-5/2$ give the main contribution to the ground
state.
\begin{figure}[btp]
\begin{center}
\vspace*{-0.5cm}
\includegraphics*[width=9.2cm]{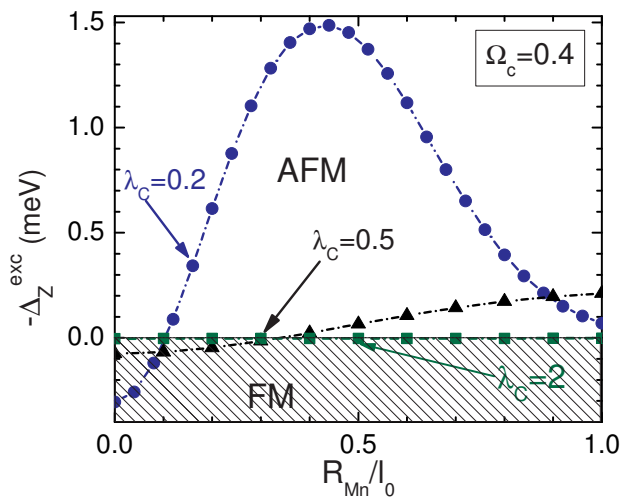}
\end{center}
\vspace{-0.5cm} \caption{ (Color online) The magnetic ion position
dependence of the exchange interaction of a three-electron quantum
dot for Coulomb strengths $\lambda_C=0.2$, $0.5$, and $2$ at
$\Omega_c=0.4$. The horizontal dotted line at $\Delta_Z^{exc}=0$
separates the plane into ferromagnetic (FM) and antiferromagnetic
(AFM) regions.}
 \label{exchangePOS_3e_severallamb}
\end{figure}

Fig.~\ref{Zeeman3e0.5l0ANDLzMzSz}(b) shows the average of the three
quantities $M_z$, $S_z$, and $L_z$ of the three-electron quantum dot
as a function of the magnetic field for a Coulomb interaction
strength $\lambda_C=0.5$. We realize that with increasing
$\lambda_C$ the $<L_z>$ ($<S_z>$) exhibits jumps at smaller critical
$\Omega_c$ (compare green dash-dot-dot curve in
Fig.~\ref{localZeeman1234e}).

\subsection{Antiferromagnetic coupling}
Now we direct our attention to the very small magnetic field
behavior. There exists a very small region of the magnetic field
where the total spin of electrons and the total spin of the magnetic
ion are oriented parallel, we found this earlier in
Ref.\cite{NgaTTNguyen} for the zero magnetic field case. These
results are now extended to nonzero magnetic field. This is made
more clear in the upper inset of Fig.~\ref{twoterms_3e_0.5l0} where
the crossing point of the two terms: the Zeeman energy of the
magnetic ion ($\Delta_{z}^{Mn}$) and the exchange interaction
($\Delta^{exc}_{z}$), occurs at $\Omega_c=0.01$ (converted to
$B\approx0.1 T$ for the considered system). This ferromagnetic
coupling extends further, up to $\Omega_c=0.04$ (see the lower inset
of Fig.~\ref{twoterms_3e_0.5l0}).
\begin{figure}[btp]
\begin{center}
\vspace*{-0.5cm}
\includegraphics*[width=9.2cm]{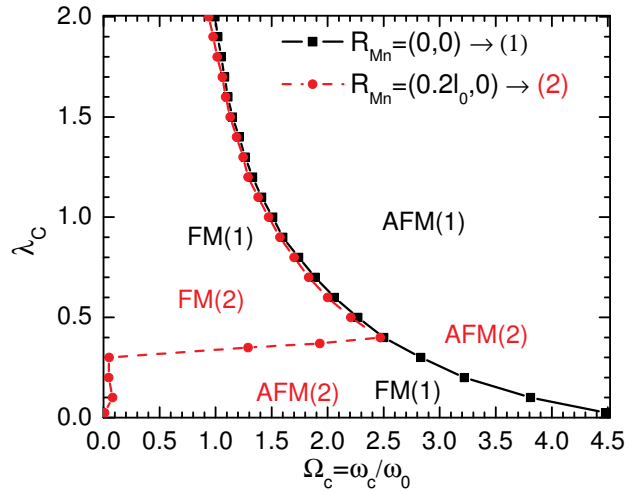}
\end{center}
\vspace{-0.5cm} \caption{ (Color online) Phase diagram for the
ferromagnetic-antiferromagnetic transition of a three-electron
quantum dot with the magnetic ion located at $(0,0)$ (black squares
referred to (1)) and ($2l_0=5.29{\AA},0$) (red circles referred to
(2)).}
 \label{PDlamb-Omegac}
\end{figure}

It is worth noticing that this ferromagnetic coupling is extended to
a much larger magnetic field range (up to $\Omega_c=2.3$) if we move
the magnetic ion to the center of the quantum dot (see
Fig.~\ref{exchange_3e_POS}). This can be understood as follows. When
the magnetic ion is located at the center of the dot and the
magnetic field is very small the absolute value of $\Delta_{z}^{Mn}$
always dominates over $\Delta^{exc}_{z}$. This is opposite to the
case when the magnetic ion is located at $(0.5l_0,0)$. Recall that
in Ref.\cite{NgaTTNguyen} we found for zero field that the exchange
Zeeman energy is minimum when the magnetic ion is at the center of
the quantum dot and approximately zero at positions very far from
the center of the quantum dot.
\begin{figure}[btp]
\begin{center}
\vspace*{-0.5cm}
\includegraphics*[width=9.2cm]{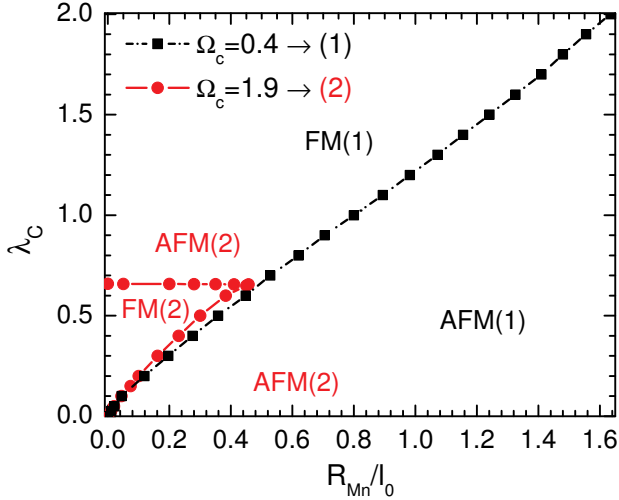}
\end{center}
\vspace{-0.5cm} \caption{ (Color online) Phase diagram for
ferro-antiferromagnetic transition of a three-electron quantum dot
in an applied field of $\Omega_c=0.4$ (black squares referred to
(1)) and $\Omega_c=1.9$ (red circles referred to (2)). }
 \label{PDlamb-RMn}
\end{figure}

Fig.~\ref{exchange_3e_POS} tells us that the magnetic field where
the antiferromagnetic coupling between electron and the magnetic ion
starts depends on the position of the magnetic ion in the quantum
dot. The system with the magnetic impurity located at $(l_0,0)$
exhibits an antiferromagnetic coupling for $\Omega_c\geq0.1$, that
is larger than $\Omega_c=0.05$ in case of $(0.5l_0,0)$.

We have discussed the appearance of antiferromagnetism in a
three-electron quantum dot. Now we go back to the two simpler cases
with the number of electrons $N_e=1, 2$ (see Fig.~\ref{exc1e2e}).
Let us first discuss the results for $N_e=1$ as given in
Fig.~\ref{exc1e2e}(a). The antiferromagnetic coupling between the
electron and the magnetic ion starts at smaller magnetic field as
the magnetic ion is moved. This is different from the previous
results for $N_e=3$. The reason is as follows: for the quantum dot
with a single electron, the electron tends to accommodate
permanently the $s$-shell with $<L_z>=0$ (see the inset of
Fig.~\ref{exc1e2e}(a)) in the ground state while the exchange
parameter in the $s$ shell ($J_{ss}$) is found to be maximum right
at the center of the quantum dot. Moving the magnetic ion away from
the center of the dot, this $J_{ss}$ is found to be smaller and as a
consequence the exchange electron-$Mn$ interaction becomes smaller
than the electron Zeeman energy, leading to an antiferromagnetic
coupling at smaller magnetic field.

The story for $N_e=2$ electrons (see Fig.~\ref{exc1e2e}(b)) is now
interesting since the two electrons accommodate the $s$ shell with
spins antiparallel making the total spin of the electron zero in the
ground state with almost unit probability. This leads to zero
contribution to the first term written in the last line in
[\ref{e:secondquantized}] for diagonal elements. Therefore, the main
contribution (even very small) to the exchange energy is now
expected to come from coupling with configurations where one of the
electrons (spin down) stays in the $s$ level and the other occupies
higher level (see the schematic diagram in Fig.~\ref{exc1e2e}(b)).
In this diagram, the magnetic ion is assumed to be located at the
center of the dot with spin down ($-5/2$). The coupling of the
electron (spin up) in the $s$ orbital with an electron from either
of the $p$ shell is zero. The only non-zero coupling is with an
electron with the quantum numbers (1,0) of the $d$ shell (as shown
in the diagram and this quantum state would change if the ion is
located away from the center of the dot) with the amount of about
$-10^{-2}$. This picture remains valid until the magnetic field is
high enough to excite one electron from the $s$ shell to a higher
quantum state forming the ground state with two up spins
antiferromagnetically coupling with the magnetic ion. For smaller
Coulomb interaction strength, the antiferromagnetic behavior occurs
at larger magnetic field since the two electrons repel each other
less and consequently they stay longer antiparallel in the $s$
shell.
\begin{figure}[btp]
\begin{center}
\vspace*{-0.5cm}
\includegraphics*[width=9.0cm]{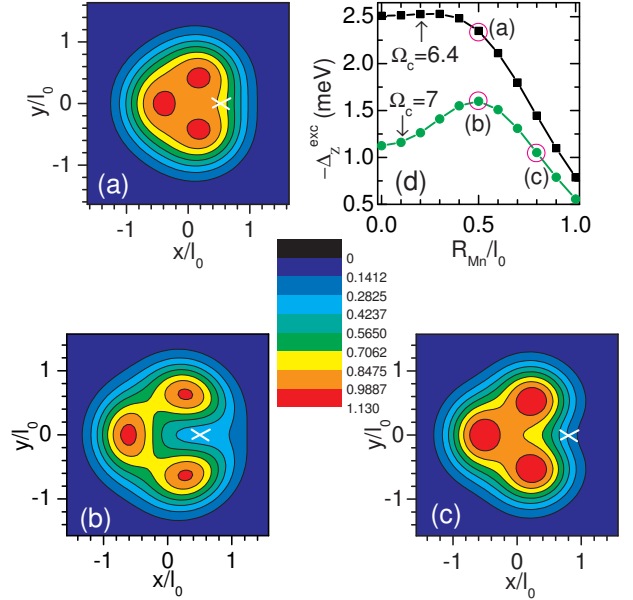}
\end{center}
\vspace{-0.5cm} \caption{ (Color online) Radial density scaled by
$l_0^2$ calculated for a three-electron quantum dot with the
magnetic ion (white cross) located at $(0.5l_0,0)$ in (a), (b) and
$(0.8l_0,0)$ in (c) when the Coulomb strength $\lambda_C=0.5$ in a
magnetic field $\Omega_c=6.4$ (see (a)) and $\Omega_c=7$ (see (b)
and (c)). (d) is the magnetic ion position dependence of
$-\Delta_Z^{exc}$ with the magenta circles indicating the respective
calculated positions of the magnetic ion in (a), (b), and (c). Red
color region is for highest value of radial density.}
 \label{DENlambda1}
\end{figure}

From the $N_e=2$ result we may ask a question: whether a
four-electron quantum dot has similar properties since the number of
electrons in both cases are even and one may have a situation where
the total spin of the electrons is zero. Indeed, if the magnetic ion
is located at the center of the quantum dot even though for $N_e=4$
the outer shell is half-filled, this is possible as illustrated in
Fig.~\ref{exc4e} where the antiferromagnetic coupling occurs at
$\Omega_c=3.87$ at which the total spin of the electrons reaches the
maximum value $<S_z>=2$. In this case, the first two electrons will
occupy the $s$ shell and the remaining two will occupy two of the
five orbitals of the $p$ and $d$ shells. This picture holds at small
magnetic field. However, there is a big difference in the exchange
energy as compared to the previous case of $N_e=2$ when the ion is
shifted away from the center of the dot, e.g. in this plot at
$(0.5l_0,0)$. The exchange energy is much larger than the result
obtained for $R_{Mn}=(0,0)$ because when the ion is out of the
center of the quantum dot, the two remaining electrons at higher
orbitals have a non-zero contribution in the diagonal exchange
elements dominating the exchange energy of the ground state. This is
the reason why the antiferromagnetic transition occurs at smaller
magnetic field ($\Omega_c=2.5$) as compared to the case when the ion
is at the center of the dot ($\Omega_c=3.87$) although the pictures
of the $M_z$ and $L_z$ transition in these cases are similar.

To complete the picture for few-electron quantum dot system, we will
discuss the AFM behavior for the system with the highest number of
electrons, $N_e=5$, where we were able to obtain accurate numerical
results. We focus on the small magnetic field region (see
Fig.~\ref{exc5e}). For the magnetic ion in the center of the dot,
the FM coupling is dominant in the shown magnetic field region
because the diagonal exchange matrix elements dominate over the
Zeeman energies of the electrons and of the magnetic ion. This is
different for the cases with the magnetic ion displaced a bit from
the center of the quantum dot. The FM-AFM occurs at $\Omega_c=0.07$
and $\Omega_c=0.12$ for $R_{Mn}=(0.5l_0,0)$ and $R_{Mn}=(l_0,0)$,
respectively. It is similar to the cases for the system with $N_e=1,
3$ due to the zero coupling between the orbitals from the $p, d$
shell with the $s$ orbital. To observe the AFM behavior for the
system with the magnetic ion located at the center of the dot where
the diagonal exchange elements are almost zero, it is crucial to
include enough quantum orbitals (that rapidly increases the size of
the Hamiltonian matrix resulting in very time consuming
calculations) so that one allows the electrons to jump to higher
energy levels and having parallel spins as previously shown for the
case $N_e=4$ (see Fig.~\ref{exc4e}). In that case, the four-electron
system exhibits an anti-ferromagnetic coupling with the magnetic ion
at the magnetic field where the total $z$-projection of the spin is
maximum $S_z=2$. The system is strongly polarized. For the case
$N_e=5$, up to $\Omega_c=0.2$, the total $S_z=0.5$ and the total
$L_z=1$. The inset in Fig.~\ref{exc5e} supports the AFM behavior for
the out of center $Mn$ as obtained in the main plot.

\subsection{Phase diagram for the ferromagnetic-antiferromagnetic transition}
Now we change the Coulomb interaction strength and explore the
magnetic behavior as function of the position of the magnetic ion.
From Fig.~\ref{exchangePOS_3e_severallamb}, it is clear that when
reducing the Coulomb interaction the system undergoes a
ferromagnetic to antiferromagnetic transition at gradually larger
magnetic fields for $Mn^{2+}$ positions that are closer to the
center of the dot. We see that $-\Delta_{Z}^{exc}$ has a peak
structure with a maximum at some specific position of the magnetic
ion, e.g. see the peak for the case $\lambda_C=0.2$ (the blue full
circles). However, it is certain that at high magnetic field, the
system is always antiferromagnetic.

The FM-AFM phase diagram for a three-electron quantum dot in
($\lambda_C,\Omega_c$) space is shown in Fig.~\ref{PDlamb-Omegac}
for two different positions of the magnetic ion. When the magnetic
ion is in the center of the quantum dot (black curve with squares in
Fig.~\ref{PDlamb-Omegac}) the critical magnetic field increases as
the Coulomb interaction strength decreases. The reason is that
increasing the Coulomb interaction helps the electrons to approach
closer the magnetic ion and therefore the critical magnetic field
for the system to transit to the antiferromagnetic phase decreases.

\begin{figure}[btp]
\begin{center}
\vspace*{-0.5cm}
\includegraphics*[width=7.8cm]{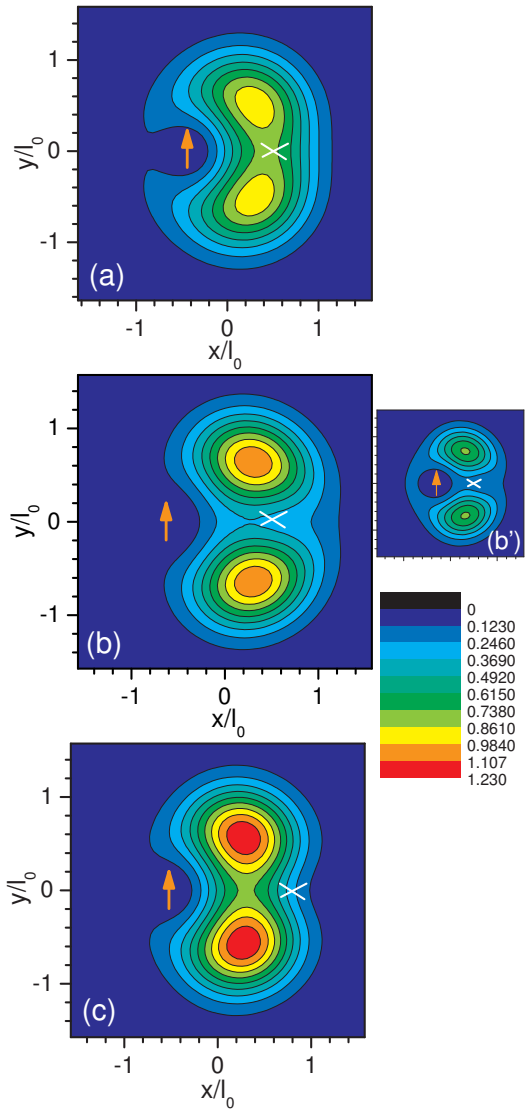}
\end{center}
\vspace{-0.5cm} \caption{ (Color online) Spin up-up pair correlation
functions scaled by $l_0^4$ calculated for the quantum dot of
Fig.~\ref{DENlambda1} where one electron (orange arrow) is pinned at
one of the three most-probable positions of the electrons as
obtained in Fig.~\ref{DENlambda1}. It is at $(-0.44l_0,0)$ in (a)
for Fig.~\ref{DENlambda1}(a), $(-0.63l_0,0)$ in (b) for
Fig.~\ref{DENlambda1}(b), and $(-0.52l_0,0)$ in (c) for
Fig.~\ref{DENlambda1}(c), respectively. (b') The same correlation
function for the case the position of the fixed electron is closer
to the ion (at $(-0.3l_0,0)$) as compared to (b). The position of
the magnetic ion is indicated by the white cross.}
 \label{paircorrPOS}
\end{figure}
Now we move the magnetic ion away from the center of the quantum dot
and we obtain the phase diagram as shown by the red curve (with
solid circles) in Fig.~\ref{PDlamb-Omegac}. For $\lambda_C<0.4$, the
stability of the FM phase with respect to an applied magnetic field
is strongly reduced and a small magnetic field turns the
three-electron system into the AFM phase. Notice that for sufficient
strong electron-electron interaction (i.e. $\lambda_C\geq0.4$) we
obtain practically the same FM-AFM phase diagram as for the case the
$Mn$-ion is located in the center of the quantum dot. A remarkable
re-entrant behavior is found in the region $0.3<\lambda_C<2$ and
$0.9<\Omega_c<2.5$ where with increasing $\lambda_C$ we go from an
antiferromagnetic to ferromagnetic and back to antiferromagnetic
phase. This unusual behavior is understood as follows. As the
impurity is moved away from the center of the quantum dot the
exchange matrix will have many nonzero off-diagonal terms that leads
to a smaller FM-AFM critical transition magnetic field. Now let us
turn our attention to the region $\lambda_C<0.4$. For very small
Coulomb interaction strength the electrons will repel each other
only weakly and are therefore pulled towards the magnetic ion (the
nonzero exchange matrix elements increase strongly) resulting in a
very small FM-AFM magnetic field. For $\lambda_C\geq0.4$, the
electrons become more strongly correlated and the critical field
stays about $\Delta\Omega_c\approx0.02\div0.07$ from the result for
$R_{Mn}=0$. If one moves the ion further and further away from the
center, the $\lambda_C\sim0.4$ transition line moves to larger
$\lambda_C$ values. For example, for $\lambda_C=0.5$ and the
magnetic ion located at $(0.5l_0,0)$ the FM-AFM critical transition
occurs at $\Omega_c=0.08$, which is much smaller than $2.21$ found
for $R_{Mn}=(0.2l_0,0)$.
\begin{figure}[btp]
\begin{center}
\vspace*{-0.5cm}
\includegraphics*[width=9.5cm]{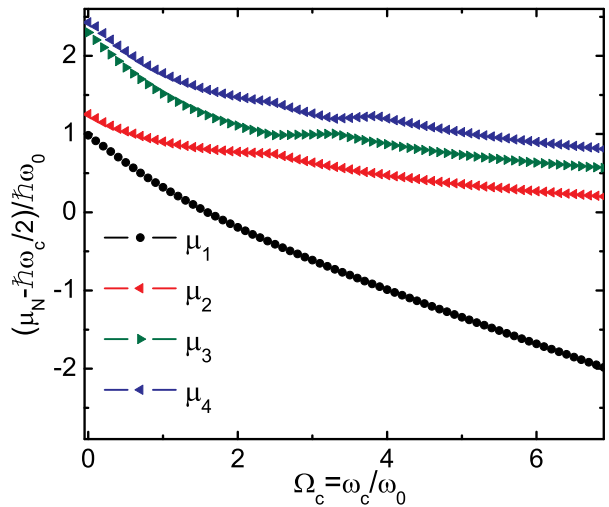}
\end{center}
\vspace{-0.5cm} \caption{ (Color online) Addition energy for
different number of electrons. The magnetic ion is located at
$(0.41l_0,0)$ and the Coulomb interaction strength is
$\lambda_C=0.2$.} \label{ChimicalPot}
\end{figure}

The dependence of the ferromagnetic-antiferromagnetic transition of
a three-electron quantum dot system on the position of the magnetic
ion is summarized in the phase diagram shown in
Fig.~\ref{PDlamb-RMn} for two different magnetic fields
$\Omega_c=0.4, 1.9$. We can predict that with slightly larger
(smaller) magnetic field the slope of the curve will be larger
(smaller). From Fig.~\ref{PDlamb-Omegac}, we already learned that
the FM-AFM transition magnetic field is largest for the ion in the
center of the quantum dot as is also seen in Fig.~\ref{PDlamb-RMn}.
The re-entrant behavior of the AFM phase as function of $\lambda_C$
is found for small $R_{Mn}/l_0$ values, i.e. when the $Mn$-ion is
not too far from the center of the quantum dot, in case the magnetic
field is not too small. The critical point
($R_{Mn}/l_0,\lambda_C$)=($0.457,0.656$) for $\Omega_c=1.9$ moves
down (up) with increasing (decreasing) magnetic field.

\subsection{Density and correlation}
In high magnetic field, the magnetic ion tends to attract electrons
because they are oppositely polarized. Because the exchange
interaction is small as compared to the Coulomb interaction, the
electrons and magnetic ion are arranged in such a way that the
electrons repel each other and also try to be as close to the
magnetic ion as possible. This picture holds above the FM-AFM
critical magnetic field.

To show this behavior explicitly, we studied the radial density and
the radial pair correlation functions. Their respective operators
are defined as:
\begin{equation}
\rho(\overrightarrow{r})=\sum_{i=1}^{N_e}\delta\left(\overrightarrow{r}-\overrightarrow{r_i}\right),
\end{equation}
and
\begin{equation}\label{e:paircorr}
C_{\sigma
\sigma^{'}}(\overrightarrow{r},\overrightarrow{r}^{'})=\sum_{i\neq
j}^{N_e}\delta_{\sigma\sigma_i}\delta\left(\overrightarrow{r}-\overrightarrow{r_i}\right)\delta_{\sigma^{'}\sigma_j}
\delta\left(\overrightarrow{r}^{'}-\overrightarrow{r_j}\right).
\end{equation}
We plot in Fig.~\ref{DENlambda1} the radial density of a three
electron quantum dot that is polarized in high magnetic field for
the case that the Coulomb strength is $\lambda_C=0.5$ and the
magnetic ion is located at two different positions for two magnetic
fields. The electrons and the magnetic ion are antiferromagnetically
coupled. The strength of that coupling can be seen from
Fig.~\ref{DENlambda1}(d) in which we plot the magnetic ion's
position dependence of the exchange energy at two magnetic fields
$\Omega_c=6.4,$ and $7$. Those magnetic fields are typical in the
sense that the exchange term is found to be very large
($\Omega_c=6.4$) or the correlation between the electrons very high
($\Omega_c=7$). Density plots are shown for $R_{Mn}$ at $(0.5l_0,0)$
and ($0.8l_0,0$). We observe three distinct peaks of maximum
probability. They are found at: $(-0.44l_0,0)$, $(0.22,0.44)l_0$,
and $(0.22,-0.44)l_0$ in Fig.~\ref{DENlambda1}(a); $(-0.63l_0,0)$,
$(0.26,0.63)l_0$, and $(0.26,-0.63)l_0$ in Fig.~\ref{DENlambda1}(b);
$(-0.52l_0,0)$, $(0.26,0.52)l_0$, and $(0.26,-0.52)l_0$ in
Fig.~\ref{DENlambda1}(c). These figures show clearly the interplay
effect where the three electrons on the one hand try to be close to
the magnetic ion and on the other hand repel each other via the
Coulomb potential energy. It results in the merging of the radial
density such that the higher the exchange energy the larger the
merging of the local maxima in the electron density and the smaller
the correlations. Fig.~\ref{DENlambda1}(d) gives an idea about the
variation of $-\Delta_Z^{exc}$ with the position of the magnetic ion
and reaches a maximum at $(0.5l_0,0)$ for $\Omega_c=7$. In
Fig.~\ref{DENlambda1}(c), the three electrons are less attracted to
the magnetic ion via the antiferromagnetic coupling as compared to
that in Fig.~\ref{DENlambda1}(b). This is due to the fact that the
$-\Delta_Z^{exc}$ for the case shown in Fig.~\ref{DENlambda1}(b) is
larger than that in Fig.~\ref{DENlambda1}(c). The electrons are
therefore found more correlated in the latter case presented by the
extended red region in Fig.~\ref{DENlambda1}(c). Thereby,
correlation between electrons in Fig.~\ref{DENlambda1}(c) is
expected to be the highest and in Fig.~\ref{DENlambda1}(a) the
smallest.

The position of the magnetic ion affects the ground-state property
as is made clear in Fig.~\ref{paircorrPOS}. We fix the spin state
and the position of one electron (indicated by the orange arrow) and
the position of the magnetic ion (the white cross). The magnetic
field is such that $\Omega_c=6.4$ in Fig.~\ref{paircorrPOS}(a) and
$\Omega_c=7$ in the others. It also reflects the fact that the
system in Fig.~\ref{DENlambda1}(a) exhibits the smallest correlation
as compared to the other two. This illustrates the point raised
above about the density. At the magnetic field $\Omega_c=7$, the
electrons are strongly polarized resulting in the red regions of the
up-up spin pair correlation function that tends to surround the
magnetic ion. We see that the three electrons are most likely to
localize around some specific positions defining a triangle with the
three electrons at the three vertices, while they are attracted to
the magnetic ion. When we locate one electron at a position closer
to the magnetic ion, see Fig.~\ref{paircorrPOS}(b'), the two peaks
decrease in amplitude as compared to those in
Fig.~\ref{paircorrPOS}(b).

\subsection{Addition energy}
The addition energy (often called the chemical potential) is defined
as the increase of the energy of the quantum dot system when an
electron is added: $\mu_{N_e}=E_{GS}(N_e)-E_{GS}(N_e-1)$. This
quantity can be measured experimentally and is plotted in
Fig.~\ref{ChimicalPot} as function of the magnetic field.
\begin{figure}[btp]
\begin{center}
\vspace*{-0.5cm}
\includegraphics*[width=9.2cm]{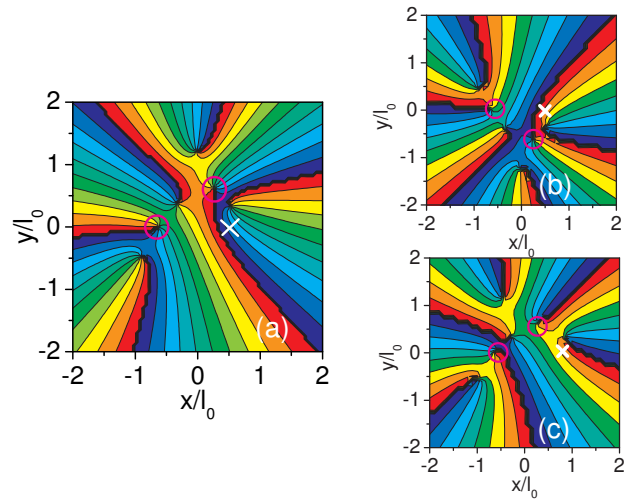}
\end{center}
\vspace{-0.5cm} \caption{ (Color online) Contour plot for the phase
of the reduced wave function of a three-electron quantum dot in
magnetic field $\Omega_c=11$ with the magnetic ion (the white cross)
located at $(0.5l_0,0)$ in (a),(b) and at $(0.8l_0,0)$ in (c) for
Coulomb strength $\lambda_C=0.5$. Two fixed electrons (indicated by
the two magenta circles) are located at the two peaks appearing in
the radial density: $(-0.63l_0,0)$ and $(0.26,0.63)l_0$ in (a), (c);
and $(-0.63l_0,0)$ and $(0.26,-0.63)l_0$ in (b). }
 \label{Vortice_lamb1}
\end{figure}

There are several cusps appearing in the addition energy curves as a
consequence of changes in the ground state. These changes are due to
variations in the z-projection of the total spin of the electrons
and/or the z-projection of the total angular momentum of the system
when the magnetic field increases beyond some specific values. The
presence of the magnetic ion leads to more cusps and the position of
these cusps is also influenced by the number of electrons and the
position of the magnetic ion. The cusps are from either of the two
systems in the study. For instance, the green triangles in
Fig.~\ref{ChimicalPot} are for $\mu_3=E_{GS}(N_e=3)-E_{GS}(N_e=2)$
has two cusps at $\Omega_c=2.6$ and $\Omega_c=3.4$. The cusp at the
point $\Omega_c=2.6$ comes from the change in the configuration of
the average of the total z-projection spin and the total
z-projection of angular momentum $(S_z,L_z)$ of the two-electrons in
the quantum dot from (0,0) to (1,1). While the other cusp
$\Omega_c=3.4$ comes from the change of the phase of the
three-electron quantum dot from $(0.5,1)$ to $(1.5,3)$. It is
similar to the case for $\mu_4$ (the blue left pointing triangles)
where the cusp appears at $\Omega_c=3.4$. At this point, we observed
a change from configuration ($0,2$) to ($1,3$). The remaining one,
$\Omega_c=3.9$, is from the four-electron case when its
configuration changes from ($1,3$) to ($2,6$).

\subsection{Vortex structure: many-body correlations}
Another way to obtain information on the correlations that are
present in the many-particle wave function is to investigate the
vortex structure. At a vortex the many-body wave function is zero
and is characterized by a change of phase of $2\pi$ when we go
around this point.
\begin{figure}[btp]
\begin{center}
\vspace*{-0.5cm}
\includegraphics*[width=8.5cm]{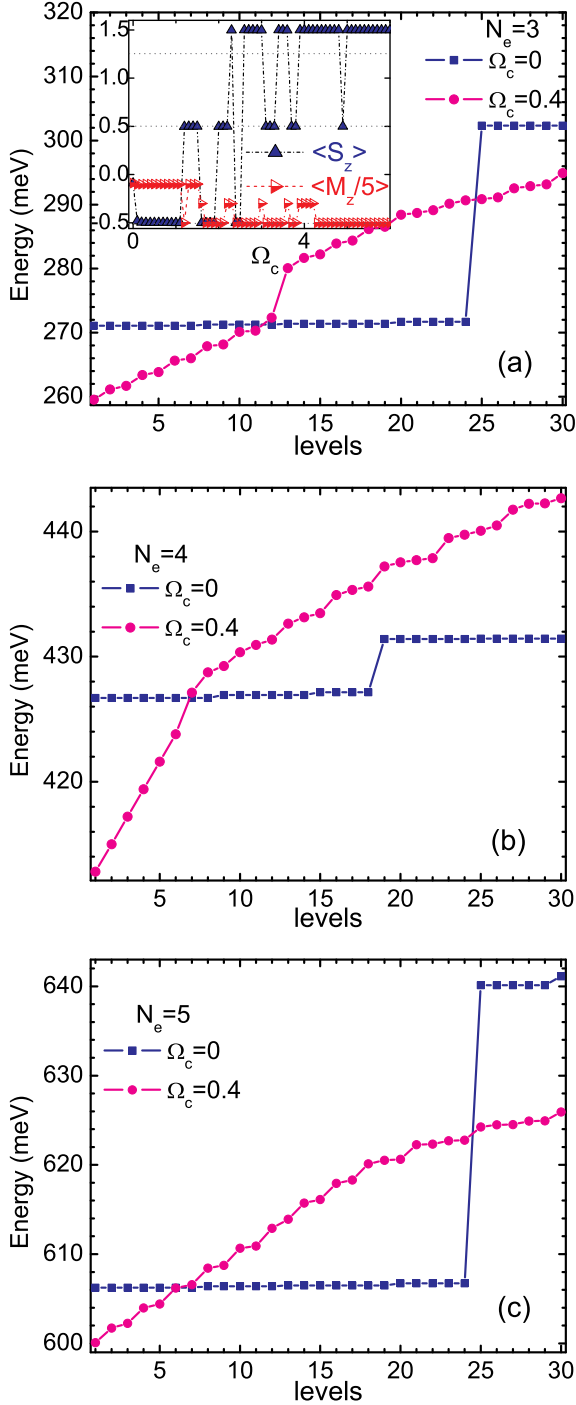}
\end{center}
\vspace{-0.5cm} \caption{ (Color online) The energy spectrum of the
first $30$ levels of the three (the top), four (the middle), and
five (the bottom) electron quantum dot with the magnetic impurity
located at $(0.5l_0,0)$. The Coulomb strength is $\lambda_C=0.5$
with magnetic fields $\Omega_c=0.0$ (the blue squares) and
$\Omega_c=0.4$ (the magenta circles). The inset in the top figure
shows the average of $M_z$ (the dark-blue triangles) and $S_z$ (the
red triangles) as a function of $\Omega_c$ of the sixth energy
level.}
 \label{spectrum3eImp0.5l0}
\end{figure}

The zeros of the wave function are similar to flux quanta when e.g.
the wave function corresponds to the order parameter in a
superconductor. The fixed electrons and the zero of the wave
function follows closely the displaced electron and one may say that
the electron plus its zero form a composite-fermion object. The
composite-fermion\cite{JKJain,Saarikoski,Marteen} (and references
therein) is a collective quasi-particle that consists of one
electron bound to an even number of vortices (flux quanta). The
composite-fermion concept introduces a new type of quasi-particle
that is used to understand the fractional quantum Hall effect in
terms of the integer quantum Hall effect of these composite
fermions.

To obtain the zeros of the wave function of the system with $N_e$
electrons, we fix $N_e-1$ electrons at some positions inside the
quantum dot and leave the remaining one free. The resulting reduced
wave function gives the probability to find the remaining electrons
at different positions in the quantum dot and the zeros' of this
function are those points where the phase of the wave function
changes by $2\pi$. As an example, we investigate the situation of a
three-electron quantum dot.
\begin{figure}[btp]
\begin{center}
\vspace*{-0.5cm}
\includegraphics*[width=9.2cm]{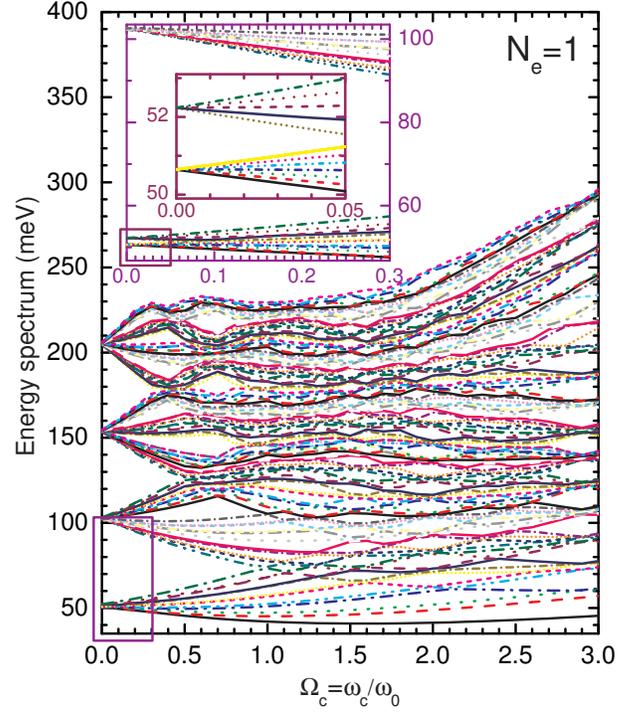}
\end{center}
\vspace{-0.5cm} \caption{ (Color online) The magnetic field
dependence of the energy spectrum (the first $120$ levels are shown)
of the one-electron quantum dot with the magnetic impurity located
at $(0.5l_0,0)$ and $\lambda_C=0.5$ (the inset is a zoom of the
first $24$ levels at low magnetic fields and the inset in the inset
shows the first $12$ levels, it is a zoom of the rectangular
region.}
 \label{spectrum-B1eImp0.5l0}
\end{figure}

Figs.~\ref{Vortice_lamb1} shows the vortex pictures of a
three-electron quantum dot containing a magnetic ion located at
positions (see the white cross) that are identical to its positions
in Fig.~\ref{DENlambda1}(a). Two, among three, electrons are fixed
at the respective peaks in the electron density. Red and black
regions are referring to the highest (2$\pi$) and lowest phase $0$,
respectively. Those plots show that there are always two vortices
near the pinned electrons' positions. For example, the number of
vortices pinned to each electron in the case $N_e=3$ at
$\Omega_c=11$ is $2$ describing the system at filling factor
$\nu=\frac{N_e(N_e-1)}{2<L_z>}\approx1/3$.  Notice also that one of
the vortices appears to be pinned at a position very close to the
$Mn$ ion.

We realize that moving the magnetic impurity to a different position
changes the relative positions of the vortices that are pinned to
the electrons with respect to one another, as shown in
Figs.~\ref{Vortice_lamb1}(a) and (c). As the electrons are
antiferromagnetically coupling to the magnetic ion this kind of
movement consequently depends on the position of the magnetic ion.
\begin{figure}[btp]
\begin{center}
\vspace*{-0.5cm}
\includegraphics*[width=9.2cm]{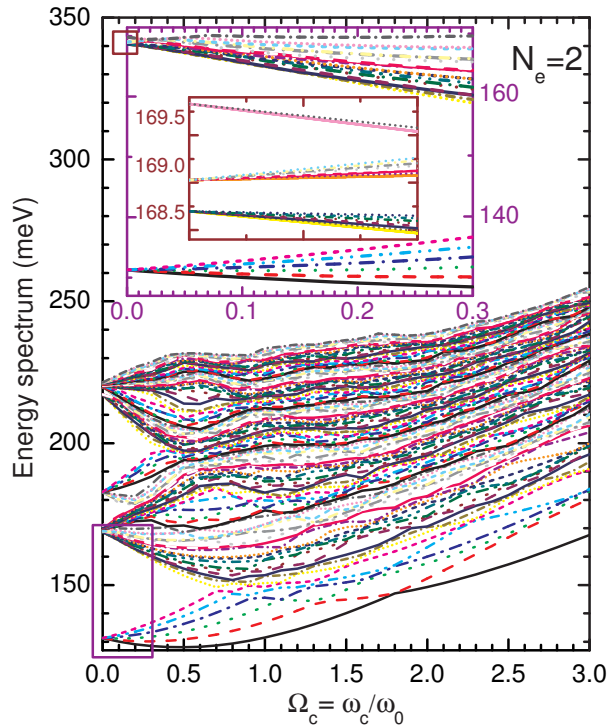}
\end{center}
\vspace{-0.5cm} \caption{ (Color online) The same as
Fig.~\ref{spectrum-B1eImp0.5l0} but now for the two-electron quantum
dot. The inset in the inset is a zoom of levels $7$th to $24$th for
magnetic field close to zero.}
 \label{spectrum-B2eImp0.5l0}
\end{figure}

In the case $\lambda_C=0.5$, we found that the average of the
maximum z-projection of the total angular momentum is $<L_z>=9$ and
the two vortices appear at the external field $\Omega_c=3.0$.
Apparently, the larger $\lambda_C$ the smaller $\Omega_c$ for which
the first two vortices appear at the pinned electrons.
\section{Energy spectrum}
In the presence of an external magnetic field, the many-fold
degeneracy of the energy spectrum of the system is lifted.
Fig.~\ref{spectrum3eImp0.5l0}(a) illustrates that point for the case
of three electrons. In the absence of the interaction between the
electrons and the magnetic ion and in the absence of a magnetic
field (blue squares), the energy spectrum is $7$-fold degenerate for
the first $7$ lowest energy levels, the next level is then $5$-fold
degenerate, and the next $7$-fold degenerate, and so on. The origin
of this was explained in Ref.\cite{NgaTTNguyen} and is due to the
coupling of the electrons and the magnetic ion. When the magnetic
field is different from zero, see red circles in
Fig.~\ref{spectrum3eImp0.5l0}, the degeneracy is lifted. In the
inset of Fig.~\ref{spectrum3eImp0.5l0} we plot $<M_z>$ (the magenta
triangles) and $<S_z>$ (the dark-blue ones) as a function of
magnetic field for the sixth level. The average of $<M_z>$ and
$<S_z>$ change abruptly as compared to those found for the
ground-state energy, e.g. see Fig.~\ref{Zeeman3e0.5l0ANDLzMzSz}.
$M_z$ and $<S_z>$ of the sixth state jump between two different
values, e.g. $-1.5$ and $-2.5$ for $<M_z>$ and $0.5$ and $1.5$ for
$<S_z>$ as function of the field. This is a consequence of
anti-crossings of energy levels as will be apparent later. The
result for four- and five-electron quantum dots are also shown in
Fig.~\ref{spectrum3eImp0.5l0}. We see the degeneracy of $8$, $6$,
$4$, and $12$ for the first $30$ levels in the case $N_e=4$ and of
$7$, $5$, $7$, $5$, $5$, and level $30$ has the same degeneracy with
the next energy level beyond the first $30$ for the case $N_e=5$ in
$B=0$ Tesla.
\begin{figure}[btp]
\begin{center}
\vspace*{-0.5cm}
\includegraphics*[width=9.2cm]{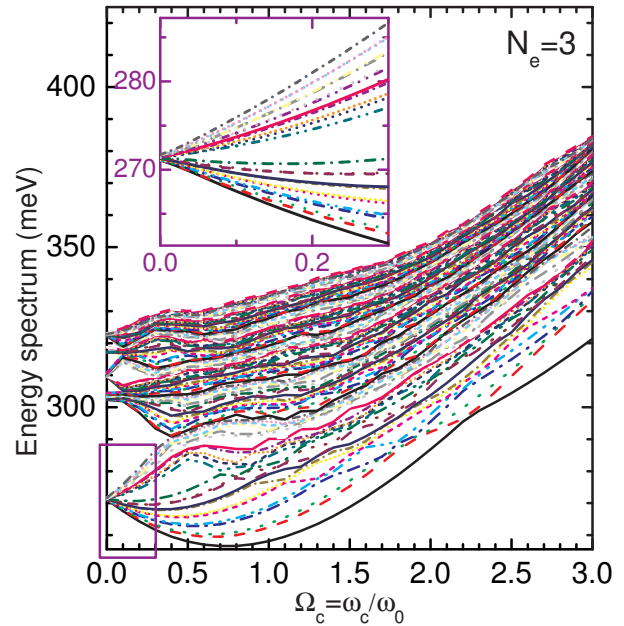}
\end{center}
\vspace{-0.5cm} \caption{ (Color online) The same as
Fig.~\ref{spectrum-B1eImp0.5l0} but now for the three-electron
quantum dot.}
 \label{spectrum-B3eImp0.5l0}
\end{figure}

To have a clearer picture of the energy spectrum of the quantum dot
system we plot in
Figs.~\ref{spectrum-B1eImp0.5l0},~\ref{spectrum-B2eImp0.5l0}, and
~\ref{spectrum-B3eImp0.5l0} the magnetic field dependence of the
first $120$ energy levels for $N_e=1, 2, 3$, respectively. The
spectra at small magnetic fields is enlarged (see insets) to see the
Zeeman splitting and the nearly-linear behavior of the energy
levels. Remember that this is due to the coupling of the electron
spins with the magnetic ion spin. For $N_e=1$ the first two levels
for $B=0$ are $7$- and $5$-fold degenerate ($7$-fold degenerate is
due to the ferromagnetic coupling of the $s$-shell electron spin
$1/2$ with the magnetic ion spin $5/2$ and $5$-fold degenerate of
that electron now with spin $-1/2$ to the magnetic ion with spin
$5/2$), respectively. A closer inspection (see
Fig.~\ref{spectrum-B1eImp0.5l0}) tells us that these $12$-levels are
exchange split into two bundles of $7$- and $5$-levels (the inset in
the inset in Fig.~\ref{spectrum-B1eImp0.5l0}). Notice that there is
a first large energy gap at very small fields between the first $12$
levels and the next $24$ ones as seen in
Fig.~\ref{spectrum-B1eImp0.5l0} while that kind of gap appears
between the first $6$ and the next $36$ for $N_e=2$
(Fig.~\ref{spectrum-B2eImp0.5l0}). For $N_e=3$
(Fig.~\ref{spectrum-B3eImp0.5l0}), this kind of gap appears after
the first $24$ energy levels. The origin is the coupling of the
third electron, which can reside at either two states of the $p$
shell while the $s$ shell is already fully filled, with the magnetic
ion with $6$ $z$-components of the spin at very small fields, i.e.
the intra-shell ($p$) exchange interaction. For $N_e=2$ the electron
ground state corresponds with a filled $s$-shell, i.e. $<S_z>=0$,
and therefor for $B=0$ only a $6$-fold degeneracy, as shown in
Fig.~\ref{spectrum-B2eImp0.5l0}, is found due to the $z$-component
of the $Mn$-spin. The next level is $8$-fold degenerate at $B=0$
($8$ come from the ferromagnetic coupling of the two-electron system
with total spin $1$ to the magnetic ion with spin $5/2$) (see the
inset in the inset in Fig.~\ref{spectrum-B2eImp0.5l0}), etc.
\begin{figure}[btp]
\begin{center}
\vspace*{-0.5cm}
\includegraphics*[width=8.7cm]{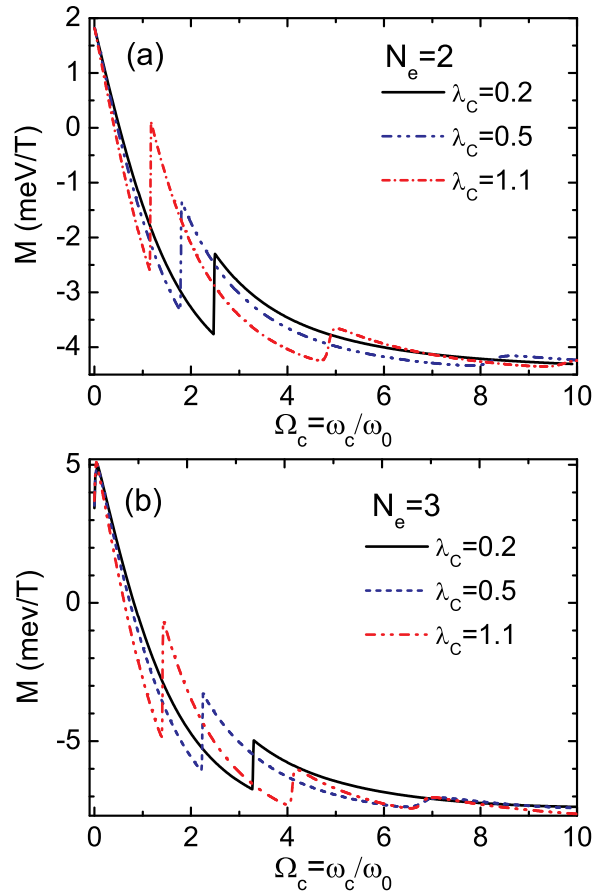}
\end{center}
\vspace{-0.5cm} \caption{ (Color online) Magnetization of the ground
state of the two- and three-electron quantum dot with the magnetic
ion located at $(0.5l_0,0)$ for three values of the Coulomb strength
$\lambda_C=0.2$, $0.5$, and $1.1$.}
 \label{Mag2and3e}
\end{figure}

With increasing magnetic field, we see that for $N_e=1$ there is
periodically an opening of energy gaps in the spectrum. Similar
energy gaps have been found earlier (as an example see e.g
Ref.\cite{Tarucha}) for a quantum dot without a magnetic impurity
and are a consequence of the electron with two-fold spin degeneracy
filling the equally-gaped-energy single-particle quantum states with
different sets of the radial and angular quantum numbers. Notice
that for $N_e=2, 3$, these gaps have disappeared.
\begin{figure}[btp]
\begin{center}
\vspace*{-0.5cm}
\includegraphics*[width=8.7cm]{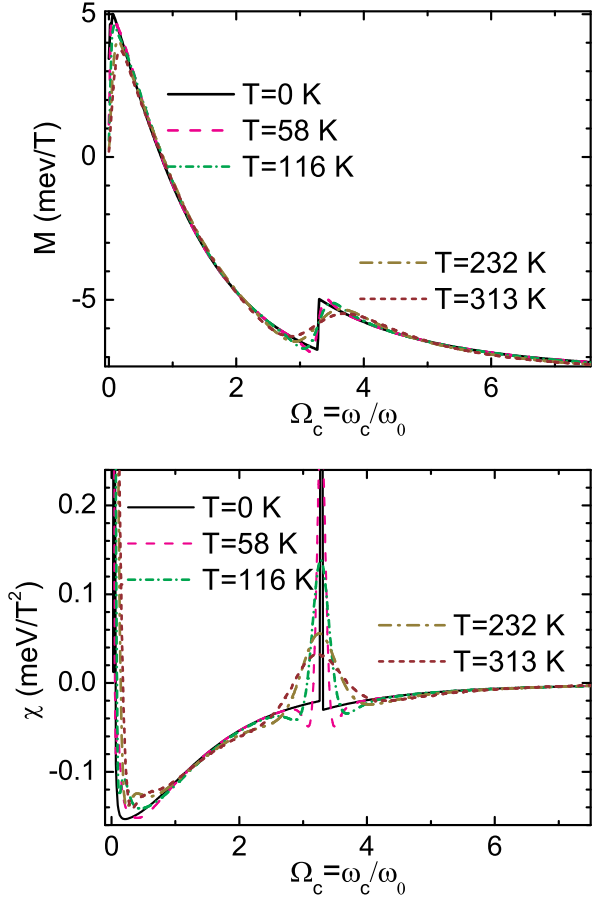}
\end{center}
\vspace{-0.5cm} \caption{ (Color online) Temperature dependence of
the magnetization and susceptibility for a three-electron quantum
dot with the magnetic ion located at $(0.5l_0,0)$ and
$\lambda_C=0.2$.}
 \label{MagTempo3e}
\end{figure}

The spectra exhibit a lot of crossings and anti-crossings, the
number of them has increased as compared to the quantum dot case
without a magnetic ion because of the Zeeman splitting of the
$Mn$-spin. When the applied field increases the gaps in the spectrum
of $N_e=1$ are still open and appear more often than in the cases of
$N_e=2, 3$. Once again, we see a lot of cusps in the energy levels
and that reminds us to abrupt changes in the configuration of the
system with magnetic field as discussed before for the ground state.
\section{Thermodynamic properties}
\subsection{Magnetization and susceptibility}
We first calculate the magnetization and susceptibility of the
system: $M=-\partial E_{GS}/\partial B$ and
$\chi=\partial{M}/\partial{B}$ at zero temperature.

The magnetization of a quantum dot with the magnetic ion located at
$(0.5l_0,0)$ with $N_e=2, 3$ electrons is plotted in
Fig.~\ref{Mag2and3e}. We see several jumps that are a consequence of
changes in the ground state, e.g. changes in $<L_z>$ (see previous
section). For example, the magnetization of the three-electron
quantum dot as plotted in Fig.~\ref{Mag2and3e}(b) for the case
$\lambda_C=0.2$ and the magnetic ion at $(0.5l_0,0)$ has a step at
$\Omega_c=3.3$. Consequently, the susceptibility also has a peak at
$\Omega_c=3.3$. The same thing happens at $\Omega_c=1.4, 4.1,$ and
$6.8$ for $\lambda_C=1.1$ in the magnetization and the
susceptibility.

For non-zero temperature, the temperature dependence of the
magnetization and susceptibility is defined by:
$M(T)=-\partial\langle E(T) \rangle/\partial B$, $\chi(T)=\partial
M(T)/\partial B$, respectively. The statistical average $<E(T)>$ is
calculated as:
\begin{eqnarray} \label{e:Cv}
\langle{E \left(\lambda,T,R_{Mn}\right)\rangle}=
\frac{\sum_{{\alpha}=1}^{N_{\alpha}}{E_{\alpha}e^{-E_{\alpha}/k_BT}}}{\sum_{{\alpha=1}}^{N_{\alpha}}{e^{-E_{\alpha}/k_BT}}},
\end{eqnarray}
where the sum is over the energy levels as displayed in e.g.
Fig.~\ref{spectrum3eImp0.5l0}.

These quantities are explored in Figs.~\ref{MagTempo3e} for $N_e=3$
and a few different temperatures (including the zero temperature
case). With increasing temperature the jumps become smoother. A very
low magnetic field peak shows up because for $T\ne0$ we have
$M\approx0$ at $\Omega_c=0$.

\subsection{Heat capacity}
An important quantity that is related to the storage of energy is
the heat capacity:
\begin{equation}
C_V\left(\lambda,T,R_{Mn}\right)=\frac{\partial{\langle{E
\left(\lambda,T,R_{Mn}\right)\rangle}}}{\partial{T}}
\end{equation}
The heat capacity is investigated as a function of the Coulomb
strength $\lambda$, temperature $T$, magnetic field, and the
position of the impurity $R_{Mn}$.
\begin{figure}[btp]
\begin{center}
\vspace*{-0.5cm}
\includegraphics*[width=8.7cm]{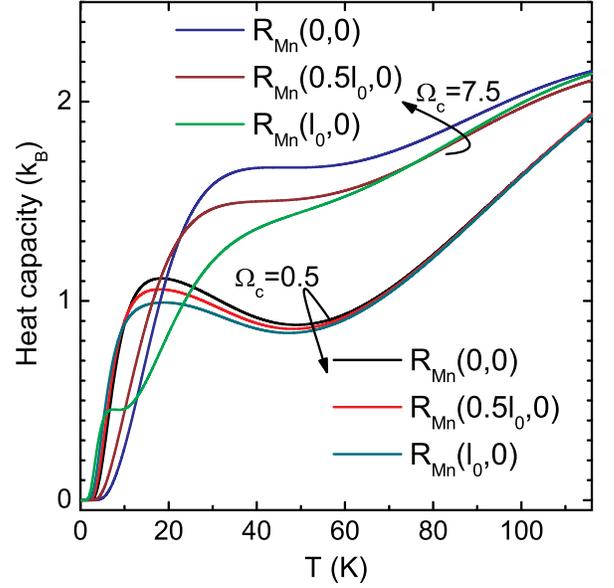}
\end{center}
\vspace{-0.5cm} \caption{ (Color online) The heat capacity vs
temperature of a three-electron quantum dot with the magnetic ion
located at three different positions, the Coulomb interaction
strength is $1.0$ and the magnetic field $\Omega_c=0.5$ (the three
lower curves) and $7.5$ (the three upper curves).}
 \label{CvseveralPOS_lamb1}
\end{figure}

We plot in Fig.~\ref{CvseveralPOS_lamb1} the specific heat for two
values of the magnetic field, i.e. $\Omega_c=0.5$ and
$\Omega_c=7.5$, and three typical positions of the magnetic ion. For
weak fields, the three electrons start to polarize and we see that
the position of the main peak moves towards higher temperature as
the magnetic ion is moved away but not too far from the center of
the dot. For the high magnetic field case the three electrons are
strongly polarized and we see a different behavior in the shift of
the main peak. This results from the change of the statistical
average of the energy levels at different fields.
\begin{figure}[btp]
\begin{center}
\vspace*{-0.5cm}
\includegraphics*[width=8.0cm]{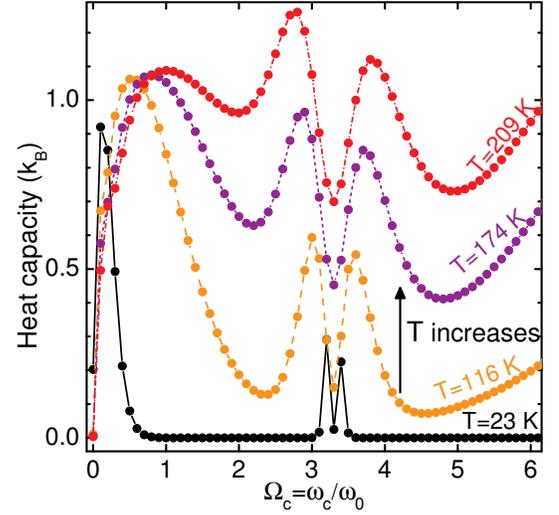}
\end{center}
\vspace{-0.5cm} \caption{ (Color online) The heat capacity vs
magnetic field of a three-electron quantum dot with the magnetic ion
located at $(0.41l_0,0)$, $\lambda_C=0.2$, at several temperatures:
$23$, $116$, $174$ and $209$ K.}
 \label{Cv-B_lamb0.4_severalT}
\end{figure}

\begin{figure}[btp]
\begin{center}
\vspace*{-0.5cm}
\includegraphics*[width=8.5cm]{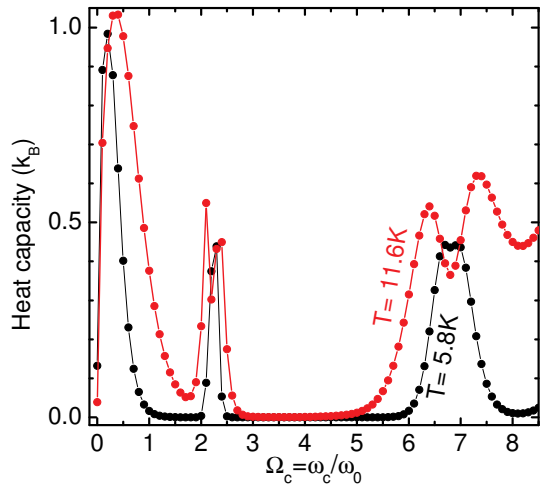}
\end{center}
\vspace{-0.5cm} \caption{ (Color online) The heat capacity vs
magnetic field of a three-electron quantum dot with the magnetic ion
located at $(0.5l_0,0)$, $\lambda_C=0.5$, and the temperatures are
$5.8 K$ and $11.6 K$.}
 \label{Cv-BTemp=5.8AND11.6}
\end{figure}

\begin{figure}[btp]
\begin{center}
\vspace*{-0.5cm}
\includegraphics*[width=8cm]{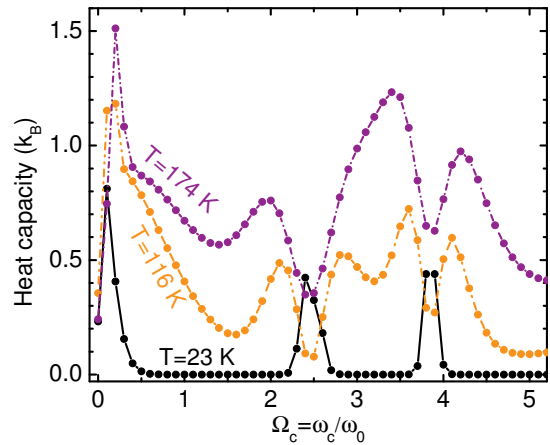}
\end{center}
\vspace{-0.5cm} \caption{ (Color online) The heat capacity vs
magnetic field of a four-electron quantum dot with the magnetic ion
located at $(0.41l_0,0)$, $\lambda_C=0.2$ plotted for several
temperatures.}
 \label{Cv-BseveralT_4e}
\end{figure}

Now we examine the behavior of the heat capacity at a specific
temperature as function of magnetic field.
Figs.~\ref{Cv-B_lamb0.4_severalT},~\ref{Cv-BTemp=5.8AND11.6},
and~\ref{Cv-BseveralT_4e} are the plots of the magnetic field
dependence of the heat capacity of three (at two Coulomb interaction
regimes) and four-electron quantum dots at some specific
temperatures and two different $\lambda_C=0.2, 0.5$. The peak at
small magnetic fields broadens and moves to higher fields with
increasing temperature. The heat capacity exhibits a number of peaks
and a clear minimum around e.g. $\omega_c=3.4$ as shown in
Figs.~\ref{Cv-B_lamb0.4_severalT}. Remember that this field
corresponds to a cusp in the energy versus magnetic field behavior
as discussed previously in subsection III.A. At very low
temperatures, this cusp still affects the heat capacity through the
sharpness of the minimum as shown in the figure and this gradually
becomes small at high temperatures. In
Fig.~\ref{Cv-BTemp=5.8AND11.6}, we see a very interesting behavior
of the heat capacity at $\Omega_c\approx2.3$: the single peak
becomes a double peak as temperature increases from $T=5.8$ to
$T=11.6 K$. This is due to the cusps now occurring around this field
in the low-energy levels of the spectrum of the three-electron
quantum dot system as observed in Fig.~\ref{spectrum-B3eImp0.5l0}.
Besides, the structure of the heat capacity is more complex (more
peaks) with increasing $\lambda_C$. This is made clear if one looks
back to the previous discussion related to
Figs.~\ref{localZeeman1234e} and ~\ref{Zeeman3e0.5l0ANDLzMzSz}(a).

For the case $N_e=4$, the heat capacity exhibits more peaks as
compared to the case $N_e=3$ and the behavior of the peaks with
increasing temperature is also very different. Temperature affects
the heat capacity of the system in the sense that it increases the
peak values and separates them in magnetic field.

The Coulomb interaction strength changes the structure of the
magnetic field dependence of the heat capacity and is illusive
through Figs.~\ref{Cv-B_lamb0.4_severalT}
and~\ref{Cv-BTemp=5.8AND11.6}. The peak of the heat capacity for the
case with smaller Coulomb interaction strength appears at higher
magnetic field as compared to the case with larger one.

\section{Discussions.}
Due to the presence of the magnetic ion (and electron-electron
interaction), electrons in the ground state do not always completely
polarize in the presence of an external magnetic field. The
configurations are mixed consisting of electrons having spins up and
down. But for very large magnetic field, the magnetic ion tends to
pull the electrons closer to the ion forming a ring-like electron
density profile. These are the consequences of the interplay of
several effects such as the Zeeman effect (on the electrons' and the
magnetic ion's spins), the Coulomb repulsion, the spin exchange
interaction, etc. This competition results in a crossover from
ferromagnetic- to antiferromagnetic coupling between the electrons
and the magnetic ion at some specific magnetic field. Interestingly,
this effect is observed to appear at higher magnetic field when we
move the magnetic ion further from the origin of the quantum dot. A
re-entrant behavior of the FM-AFM transition is found as function of
the Coulomb interaction strength when the magnetic ion is moved out
(but not too far) from the center of the quantum dot.

The energy levels exhibit cusps as function of the magnetic field
which correspond to changes of the configuration of the system as
expressed by the values of ($<S_z>,<L_z>$). These cusps move to
lower magnetic field with increasing Coulomb interaction strength.
The number of cusps increases with increasing number of electrons.
These cusps show up in the addition energy.

The transformation of the electron system to those of composite
fermions is studied. In high magnetic fields, the electrons attach
an even number of quantized vortices which we made clear by
examining the many-body ground-state wave function in the presence
of a magnetic ion. Unlike the case without a magnetic ion where all
the vortices are tightly bound to the electrons, when we fix the
electrons at different positions the system of vortices stays pinned
to the electrons and moves with the electrons but the relative
positions of the vortices are modified.

The contribution of the local Zeeman splitting energy to the total
energy of the system in large external fields is very small as
compared to the contributions from the other parts. However, a
slight movement of the position of the magnetic ion inside the
quantum dot affects the result, slightly.

With increasing applied magnetic field, each time the system jumps
to a different $(<L_z>,<S_z>)$ configuration leads to the appearance
of a peak in the thermodynamic quantities as e.g. the susceptibility
and the heat capacity. In the presence of the magnetic ion, the
structure of peaks in the heat capacity changes with the position of
the magnetic ion. As temperature increases, these peaks split into
two peaks and become smoother.

\section{Acknowledgments}
This work was supported by FWO-Vl (Flemish Science Foundation), the
EU Network of Excellence: SANDiE, and the Belgian Science Policy
(IAP).


\end{document}